\documentclass{ws-ijmpe}

\begin{document}

\markboth{Nathan Grau}{Evolution of the Away-Side Jet Shapes in
$\pi^{0}-h^{\pm}$ Correlations in 200 GeV Au+Au Collisions with
RHIC-PHENIX}

\catchline{}{}{}{}{}

\title{Evolution of the Away-Side Jet Shapes in
$\pi^{0}-h^{\pm}$ Correlations in 200 GeV Au+Au Collisions with
RHIC-PHENIX}

\author{Nathan Grau, {\it for the
PHENIX\footnote{For the full list of PHENIX authors and
acknowledgements, see Appendix 'Collaborations' of this volume}
Collaboration}}

\address{Columbia University, Nevis Labs,\\
P.~O.~Box 137 Irvington, NY, 10533, United States\\
ncgrau@nevis.columbia.edu}

\maketitle

\begin{history}
\received{(received date)}
\revised{(revised date)}
\end{history}

\begin{abstract}
Two-particle azimuthal correlations have allowed detailed study of
the modification to di-jets in the hot, dense medium created in RHIC
collisions. Light can be shed by such correlations on many novel
effects discovered at RHIC such as the enhanced $p/\pi^{+}$ ratio
and conical structure of events with energy loss. Using $\pi^{0}$
triggers to reduce the effect of recombination we utilize
high-$p_{T}$ correlations to explore the away-side structure.
\end{abstract}

\section{Introduction}
The studies of hard scattering processes in the heavy ion
environment, previously uncharted before the advent of RHIC, have
resulted in very dramatic results. Initial results, and the most
cited, were the single particle studies where a suppressed
production of high-$p_{T}$ particles compared to binary-scaled $p+p$
collisions was observed~\cite{PHENIXpi0}. Two-particle studies, via
the azimuthal correlation between two high-$p_{T}$ particles, have
extended these single particle studies to infer properties of the
jets themselves responsible for the particle production. At $p_{T}$
of 2-5 GeV/c, where hard scattering production dominates in $p+p$
collisions, there are strong modifications to the structure of the
two-particle correlations in Au+Au collisions. In both the near-side
and away-side the jets show yields which are enhanced compared to
$p+p$ at low-$p_{T,assoc}$ and suppressed at the highest
$p_{T,assoc}$~\cite{ChunQM06}. The shape of the away-side
correlations at these intermediate $p_{T}$ are also broader and show
a humped structure compared to those in $p+p$
collisions~\cite{PPG068}. Another difference in particle production
between $p+p$ and $Au+Au$ at these $p_{T}$ is the observation of
enhanced baryon production. The $p/\pi^{+}$ ratio is a factor of
$\sim$5 larger than that measured in $p+p$~\cite{PHENIXReco}. These
ratios and spectra are consistent with models of recombination of
partons from the medium and partons from jet
fragments~\cite{HwaReco}. Recent results indicate that the structure
of the away-side seems to be dependent on the flavor of the
associated particle~\cite{AnneQM06}.

This contribution focuses on $\pi^{0}-h^{\pm}$ correlations at high
trigger $p_{T}$ ($>$5 GeV/c). This is motivated for several reasons.
First, given a high-$p_{T}$ $\pi^{0}$, the effects on these
correlations from recombination is small~\cite{PPG072} and less than
the effects from using an unidentified hadron trigger. Next, it is
instructive to measure the away-side structure of these correlations
at high-$p_{T}$ to determine if the structure is similar to the
lobed structure observed in lower-$p_T$ correlations. Finally, since
the away-side parton on average travels through a longer medium path
length, effects of energy loss should be more evident on the
away-side. Whatever away-side structure exists, it is important to
measure the yield and shape to quantify this modification compared
to $p+p$.

\section{High-p$_{\rm{T}}$ Azimuthal Correlations}

Azimuthal correlations are a well-established tool to measure jet
fragments in all RHIC collisions. Details of the method used to
measure correlations are found in reference~\cite{PPG039}. Briefly,
azimuthal correlations are measured with events containing a trigger
particle. Detector acceptance and efficiency correlations are
measured by correlating particles from different events. These are
then removed from the real pair correlations to reveal the physical
correlations.

\begin{figure}[b]
\centerline{\psfig{file=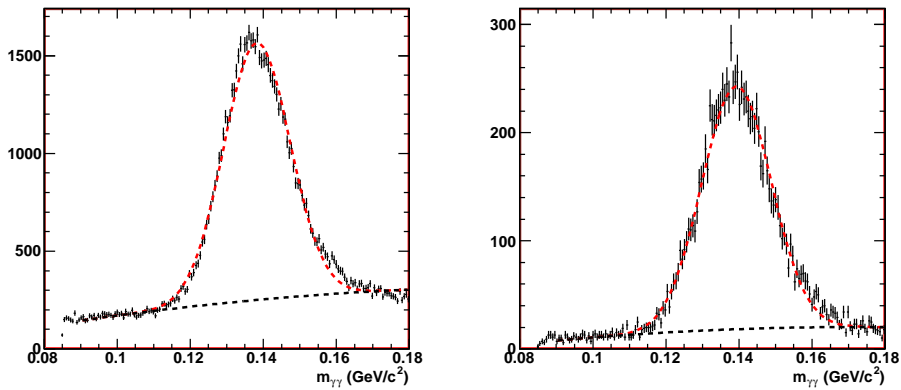,width=\textwidth}}
\vspace*{8pt} \caption{Di-photon invariant mass distributions for
0-20\% Au+Au collisions and photon pair $p_{T}$ from 5-7 GeV/c
(left) and 7-20 GeV/c (right). Dashed line is a Gaussian and
polynomial fit to describe the signal and
background.}\label{fig:pi0mass}
\end{figure}

The $\pi^{0}$ trigger particles are measured in PHENIX by their
$\gamma\gamma$ decay channel in the electromagnetic calorimeter. One
experimental advantage of using $\pi^{0}$ triggers is the ability to
measure the signal-to-background (S/B) of the triggers and to
measure the correlations due to these fake $\pi^{0}$ triggers from
combinatoric photon pairs within the $\pi^{0}$ mass. Example
di-photon invariant mass peaks are shown in Fig.~\ref{fig:pi0mass}
for 0-20\% Au+Au collisions for the two trigger ranges considered
here, 5-7 GeV/c and 7-20 GeV/c. The S/B is 3.6 for the 5-7 GeV/c and
8.6 for 7-20 GeV/c triggers. To measure the effect on the
correlations due to the fake $\pi^{0}$ triggers, correlations with
di-photon triggers in mass ranges below and above the $\pi^{0}$ peak
are measured. The correlations from these combinatoric photon
triggers within the $\pi^{0}$ mass window are then extrapolated. In
this analysis since the S/B is sufficiently large the effect of the
background correlations is 5\% nearly independent of $\Delta\phi$
and as such we assign a 5\% systematic error on the correlations.

\begin{figure}[t]
\centerline{\psfig{file=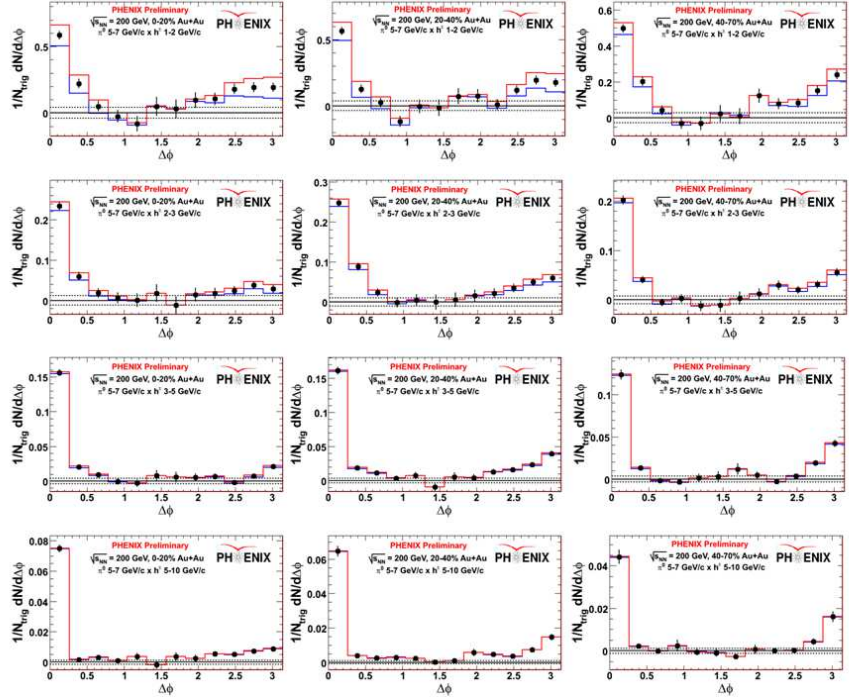,width=\textwidth}}
\vspace*{8pt} \caption{Correlations between $\pi^{0}$ triggers from
5-7 GeV/c and associated hadrons from 1-2 GeV/c (upper row), 2-3
GeV/c (second row), 3-5 GeV/c (third row), and 5-10 GeV/c (lower
row) in Au+Au collisions in 0-20\% central (left column), 20-40\%
central (middle column), and 40-70\% (right column). Statistical
errors are indicated by the error bars on the points. The systematic
error on the normalization is given as the dashed line bracketing
zero. The systematic error due to the $v_{2}$ uncertainty is the
solid histogram around the data points. Not shown is a systematic
error of 10\% due to the single hadron efficiency and 5\% from the
combinatoric contribution under the $\pi^{0}$
peak.}\label{fig:cf5-7trigg}
\end{figure}

Once the physical correlations are measured, the interest is in the
correlations due to jets. In $A+A$ collisions the elliptic flow
results in a physical two-particle correlation. This is removed by
assuming that the correlations can be decomposed into two
sources~\cite{PPG032}
\begin{equation}\label{eq:cfdecomp}
\frac{1}{N_{trig}}\frac{dN}{d\Delta\phi} =
B\left(1+2v_{2}^{trig}v_{2}^{assoc}\cos\left(2\Delta\phi\right)\right)
+ \mathcal{J}\left(\Delta\phi\right)
\end{equation}
The $v_{2}$ of the trigger and associated particles are measured
independently from an analysis of the single $\pi^{0}$ and charged
hadrons with respect to the reaction plane~\cite{pi0v2}. The
background level $B$ must be determined in order to subtract the
elliptic flow contribution. This is done by the Zero Yield At
Minimum (ZYAM) method~\cite{PPG032}. In the end the jet correlations
have errors due to background determination from ZYAM and the
uncertainties in the measured $v_{2}$ values.

\begin{figure}[t]
\centerline{\psfig{file=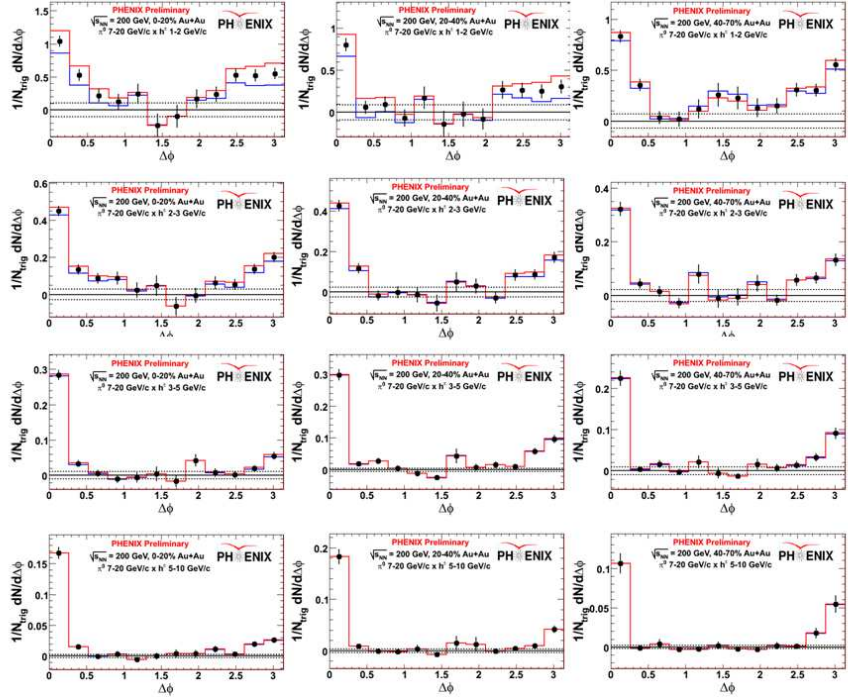,width=\textwidth}}
\vspace*{8pt} \caption{Correlations between $\pi^{0}$ triggers from
7-20 GeV/c and associated hadrons from 1-2 GeV/c (upper row), 2-3
GeV/c (second row), 3-5 GeV/c (third row), and 5-10 GeV/c (lower
row) in Au+Au collisions in 0-20\% central (left column), 20-40\%
central (middle column), and 40-70\% (right column). Statistical
errors are indicated by the error bars on the points. The systematic
error on the normalization is given as the dashed line bracketing
zero. The systematic error due to the $v_{2}$ uncertainty is the
solid histogram around the data points. Not shown is a systematic
error of 10\% due to the single hadron efficiency and 5\% from the
combinatoric contribution under the $\pi^{0}$
peak.}\label{fig:cf7-20trigg}
\end{figure}

\section{Results and Discussion}
The resulting jet correlations between $\pi^{0}$ triggers and
associated hadrons are shown in Fig.~\ref{fig:cf5-7trigg} and
Fig.~\ref{fig:cf7-20trigg} for $\pi^{0}$ triggers from 5-7 GeV/c and
7-20 GeV/c, respectively. The associated hadron $p_{T}$ varies top
to bottom in the bins 1-2 GeV/c, 2-3 GeV/c, 3-5 GeV/c and 5-10
GeV/c, respectively. The centrality of the collisions varies left to
right in bins of 0-20\%, 20-40\%, and 40-70\%, respectively. The
dashed lines on the correlations around the zero line indicate the
uncertainty in the background level from the ZYAM procedure. The
solid lines around the points indicate the error on the $v_{2}$. All
errors are 1$\sigma$.

From these jet correlations no obvious lobed structure is observed
as in lower trigger $p_T$ correlations. A statistically significant
away-side distribution is observed at all $p_T$ and centrality. What
is not clear from these data alone are any systematic trends in
centrality, associated $p_T$ , or trigger $p_T$ for yield at
$\Delta\phi\sim$2 rad.

In the future it will be necessary to compare directly to the
correlations in p+p and to quantify whatever yield exists in Au+Au
compared to p+p to search for the existence of conical structure at
these $p_{T}$. Further, it will be important to extend the $\pi^{0}$
trigger $p_{T}$ reach down to 2-5 GeV/c where the strong
modifications are seen in unidentified hadron triggered
correlations.

\end{document}